\documentclass[jgrga]{agutex2015}

 \usepackage{lineno}
  \usepackage{graphicx}
  \usepackage{color}
  \setkeys{Gin}{draft=false}
\authorrunninghead{ISHIKAWA ET AL.}

\titlerunninghead{FINE-PITCH CDTE HARD X-RAY DETECTOR}

\begin{document}

%
%

\title{Fine-pitch CdTe detector for hard X-ray imaging and spectroscopy of the Sun with the \textit{FOXSI} rocket experiment}
%
%

%
%




\author{Shin-nosuke Ishikawa, \altaffilmark{1}
Miho Katsuragawa, \altaffilmark{1,2}
Shin Watanabe, \altaffilmark{1,2}
Yuusuke Uchida,\altaffilmark{1.2}
Shin'ichiro Takeda, \altaffilmark{3}
Tadayuki Takahashi, \altaffilmark{1,2}
Shinya Saito, \altaffilmark{4}
Lindsay Glesener,\altaffilmark{5}
Juan Camilo Buitrago-Casas, \altaffilmark{6}
S\" am Krucker, \altaffilmark{6,7}
Steven Christe\altaffilmark{8}
       }

\altaffiltext{1}{Institute of Space and Astronautical Science, Japan Aerospace Exploration Agency, Sagamihara, Kanagawa, Japan}
\altaffiltext{2}{Graduate School of Science, University of Tokyo, Bunkyo, Tokyo, Japan}
\altaffiltext{3}{Advanced Medical Instrumentation Unit, Okinawa Institute of Science and Technology Graduate University, Onna, Kunigami, Okinawa, Japan}
\altaffiltext{4}{Department of Physics, Rikkyo University, Toshima, Tokyo, Japan}
\altaffiltext{5}{School of Physics and Astronomy, University of Minnesota, Minneapolis, Minnesota, USA}
\altaffiltext{6}{Space Science Laboratory, University of California, Berkeley, California, USA}
\altaffiltext{7}{Institute of 4D Technologies, School of Engineering University of Applied Sciences Northwestern Switzerland, Windisch, Aargau, Switzerland}
\altaffiltext{8}{Goddard Space Flight Center, National Aeronautics and Space Administration, Greenbelt, Maryland, USA}





%
%


\keypoints{
\item A fine-pitch CdTe detector has been developed for a sounding rocket hard X-ray observation of the Sun.
\item The required fine position resolution and low noise were achieved in laboratory calibrations.
\item An observation was successfully performed and hard X-ray images of the Sun were obtained by the detector.
}


%
%


\begin{abstract}
We have developed a fine-pitch hard X-ray (HXR) detector using a cadmium telluride (CdTe) semiconductor for imaging and spectroscopy for the second launch of the Focusing Optics Solar X-ray Imager (\textit{FOXSI}).  
\textit{FOXSI} is a rocket experiment to perform high sensitivity HXR observations from 4--15 keV using the new technique of HXR focusing optics.  The focal plane detector requires $<$100~$\mu$m position resolution (to take advantage of the angular resolution of the optics) and $\approx$1 keV energy resolution (FWHM) for spectroscopy down to 4 keV, with moderate cooling ($> -30$~$^{\circ}$C).  Double-sided silicon strip detectors were used for the first \textit{FOXSI} flight in 2012 to meet these criteria.  To improve the detectors' efficiency (66\,\% at 15 keV for the silicon detectors) and position resolution of 75~$\mu$m for the second launch, we fabricated double-sided CdTe strip detectors with a position resolution of 60~$\mu$m and almost 100\,\% efficiency for the \textit{FOXSI} energy range.  The sensitive area is 7.67~mm $\times$ 7.67~mm, corresponding to the field of view of 791'' $\times$ 791''.  An energy resolution of 1 keV (FWHM) and low energy threshold of $\approx$4 keV were achieved in laboratory calibrations.  The second launch of \textit{FOXSI} was performed on December 11, 2014, and images from the Sun were successfully obtained with the CdTe detector.  Therefore we successfully demonstrated the detector concept and the usefulness of this technique for future HXR observations of the Sun. 
\end{abstract}

%
%

%

\begin{article}

%
%

\section{Introduction}
The Sun emits hard X-rays (HXRs, here defined as X-rays above a few keV) from 
high temperature ($>$10~MK) plasma and from accelerated, non-thermal particles in the solar atmosphere.  
HXR observations of the Sun are an important tool to investigate energy release, especially in solar flares, since flares accelerate particles to high energies and heat ambient plasma.
From the first solar HXR observation to the present day, the standard observational techniques are indirect, such as the modulation collimators onboard \textit{Yohkoh} \citep{ogawara1991, kosugi1991} and the Reuven Ramaty High-Energy Solar Spectroscopic Imager (\textit{RHESSI}, \citet{lin2002}) spacecraft.
The Spectrometer / Telescope for Imaging X-rays (\textit{STIX}) onboard the \textit{Solar Orbiter} mission to be launched in 2017 also uses the modulation collimator because of the limitations of weight and space \citep{krucker2013}.  
The key to understanding flare particle acceleration likely lies in detailed observation of coronal HXR sources above magnetic loops \citep{masuda1994, ishikawa2011lt, krucker2014lt}, as it is here the particles are thought (in some, but not all theories) to be accelerated. 
However, these coronal sources are much fainter than sources at loop footpoints in general \citep{krucker2008}, and a high imaging dynamic range is required to study these faint and bright sources simultaneously.  The necessary imaging dynamic range is not available with indirect imagers.

Furthermore, high sensitivity is necessary for further solar HXR study of non-flaring regions.  The Fourier imaging of \textit{RHESSI}, for example, requires on the order of 1000 photons to produce a good image.
Faint, well-distributed sources do not modulate as well as bright, compact ones.  Because of the limited detection capabilities of indirect imaging, no significant HXR source has ever been imaged in the quiet Sun (outside active regions) \citep{hannah2010}.  

Thanks to recent technology improvements, direct HXR imaging using reflecting optics is now possible, and HXR measurement no longer needs to rely on indirect methods.  In addition to improved effective area, direct imagers achieve a higher signal to noise ratio by focusing HXRs to a small detector area (with correspondingly lower backgrounds), resulting in a high sensitivity. 
The \textit{NuSTAR} spacecraft (\citet{harrison2013}, launched in 2012) and Hard X-ray Imager (HXI) \citep{sato2016, sato2014} onboard the \textit{Hitomi} (formerly called \textit{ASTRO-H}) spacecraft (\citet{takahashi2014} , launched in February 2016) have already begun astrophysical observations 
using HXR focusing optics.  \textit{NuSTAR} also performs a few solar observations per year, though it is not optimized for the bright fluxes of the Sun \citet{grefenstette2016, hannah2016}.
The sounding rocket experiment Focusing Optics X-ray Solar Imager (\textit{FOXSI}) is a mission to adapt and optimize direct focusing HXR optics for solar purposes.  
The \textit{FOXSI} optics are Wolter-I grazing incidence telescopes developed in NASA Marshall Space Flight Center with an angular resolution of 4.3$\pm$0.6~arcseconds (FWHM) on-axis \citep{krucker2014}.  
The focal length is limited by the payload size to 2~m, which restricts the energy range to 4--15~keV because of the energy dependence of reflection angles. 
Since the observation time on a sounding rocket payload is limited to $\approx$6~minutes, large solar flares such as X- or M-class flares in the X-ray flux classification by the Geostationary Operational Environmental Satellite (\textit{GOES}) cannot be expected.  
Instead, the observational targets of \textit{FOXSI} are non-flaring regions such as quiescent active regions and the quiet Sun.

The first \textit{FOXSI} flight on November 2, 2012 produced the first focused image of the Sun above 5 keV: a partly occulted microflare.  \textit{FOXSI}'s high dynamic range was successfully demonstrated compared to the observation of the same flare by \textit{RHESSI} \citep{krucker2014}.
Although a non-flaring active region lay within one of the \textit{FOXSI} targets, no significant HXR emission was detected.  The result strongly constrained the amount of high temperature ($>$8~MK) plasma in this region \citep{ishikawa2014}.

To make full use of the \textit{FOXSI} rocket's sensitivity and resolution places the following requirements on the detector:
\begin{enumerate}
\item Efficiency in the \textit{FOXSI} energy range up to 15~keV.
\item Position resolution comparable to the optics resolution of $\approx$40~$\mu$m.  
\item Low noise for spectroscopy down to 4~keV under the moderate cooling easily achievable in a rocket payload ($\approx -30~^{\circ}$C at lowest) with no active cooler in flight.
\item Count rates of up to a few hundred counts per second so as to enable high livetime even with the high count rates expected from an active region.  
\end{enumerate}
To meet those requirements, we developed a Double-sided Silicon Strip Detector (DSSD) for the first flight of \textit{FOXSI} \citep{ishikawa2011, saito2010}.
A double-sided strip detector (also called a cross strip detector) has strip electrodes on both the top and bottom surfaces.  
The strip electrodes on each surface are parallel, and top strips and bottom strips are placed orthogonally. 
By reading out both top and bottom electrodes, $x$ and $y$ positions of an incident photon can be identified, enabling 2 dimensional imaging.
The \textit{FOXSI} DSSD has a position resolution of 75~$\mu$m and efficiency of 66\% at 15~keV.  
It has 128 strips on each surface, and positional information corresponds to $128 \times 128 = 16384$~pixels are obtained by reading out only $128+128=256$~channels.  It is convenient to use this small number of channels to operate the HXR imaging detectors in the limited power and computational resources of the rocket payload.  

To achieve better efficiency for $>$10~keV, cadmium telluride (CdTe) is a desirable material because of the high atomic numbers and high density.  
The CdTe double-sided strip detectors (also called CdTe-DSDs in some papers, \citet{watanabe2009, ishikawa2010}) we have developed for \textit{Hitomi}/HXI are good candidates for the improved \textit{FOXSI} detector.  
For the second flight (\textit{FOXSI}-2), we developed a CdTe double-sided detector with finer pitch than the \textit{FOXSI} DSSD.  
In this article, we summarize the specification, evaluation results and in-flight performance of the \textit{FOXSI} CdTe detector.  

\section{Detector Design} 
\begin{figure}[htbp]
 \begin{center}
 \noindent\includegraphics[width=18pc]{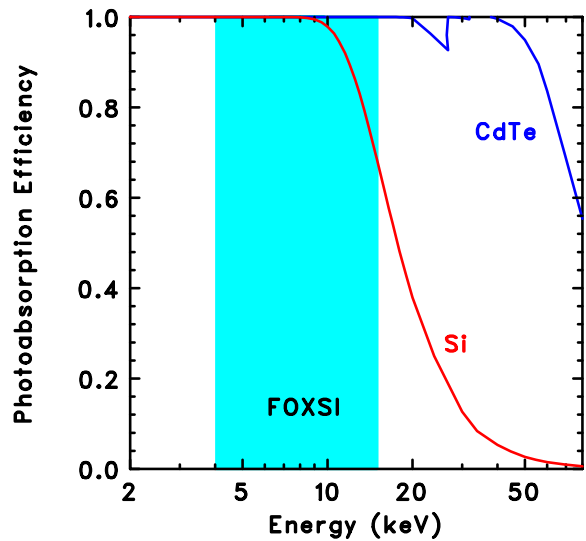}
 \caption{Photoabsorption efficiency of 0.5~mm thick silicon and CdTe detectors.  \textit{FOXSI} energy range is highlighted.}
 \label{figure_efficiency}
 \end{center}
 \end{figure}

We developed the \textit{FOXSI} CdTe double-sided strip detector with a thickness of 0.5~mm.  
The detection efficiency of 0.5~mm thick silicon (\textit{FOXSI} DSSD) and CdTe (\textit{FOXSI}  CdTe) are compared in Fig.~\ref{figure_efficiency}.
While the efficiency of the silicon detector drops below 100\% for $>$10~keV, the CdTe detector has almost 100~\% for the entire \textit{FOXSI} energy range.  
The \textit{FOXSI} CdTe detector also demonstrates its usefulness for a future solar HXR instrument with a wider energy range; it has $>$50\% efficiency up to 80~keV.

The strip size of the \textit{FOXSI} CdTe detector is 7.67~mm $\times$ 0.05~mm, and the strip pitch is as fine as 60~$\mu$m with a 10~$\mu$m gap between strips.
128 strips are placed on each of the top and bottom surfaces and the sensitive area is 7.67~mm $\times$ 7.67~mm.  
This corresponds to a field of view of 791'' $\times$ 791'' (13 arcmin) at the \textit{FOXSI} rocket focal length of 2~m. 
As with the \textit{FOXSI} DSSD, a 16384~pixel position resolution can be obtained by reading out only 256~channels.  
The CdTe sensors were fabricated by the ACRORAD company in Okinawa, Japan.  
To improve the quality of the sensor, the traveling heater method (THM) is used for the crystal growth \citep{funaki1999}. 
The material of the electrodes are aluminum for one surface, and the platinum for the other. 
A Schottky diode structure is formed by the contact of the aluminum electrodes and CdTe, and the contact between the platinum electrode and CdTe is Ohmic.  
By applying a bias voltage between the aluminum electrodes (High Voltage side, or HV side) and the platinum electrodes (Low Voltage side, or LV side), 
the leakage current is significantly reduced even with a bias voltage of only a few hundred volts. 
The mobility-lifetime products of the charge carriers in CdTe is lower than those of silicon\citep{eisen1998}.
CdTe detectors with good energy resolutions are successfully developed using the combination of the THM crystal growth and Schottky diode formation by our group \citep{takahashi2001, takahashi1999}.  
By using aluminum as the HV electrodes, pixelated or strip electrode formation become available \citep{watanabe2007, ishikawa2007} and Schottky CdTe double-sided strip detectors were successfully developed \citep{watanabe2009, ishikawa2010}.
Guard-ring electrodes are placed around the strips for both anode and cathode surfaces, and those are connected to the voltage levels of the HV and LV sides. 
A significant fraction of leakage current passes through the side surface of a CdTe detector, and the guard-ring electrode reduces leakage current in the imaging area \citep{nakazawa2004}.
With a guard-ring width of 0.5~mm and an edge-strip-to-guard-ring gap of 0.05~mm, the physical size of the detector is 8.87~mm $\times$ 8.87~mm.

Low-noise Application Specified Integrated Circuits (ASICs) known as VATA450 are used to read out the signals from the detector.  
The VATA450 was designed for the Soft Gamma-ray Detector (SGD) \citep{fukazawa2014, watanabe2014} onboard \textit{Hitomi} by our group in collaboration with IDEAS in Norway.  
While the function is the same as the VATA451 ASIC used for the \textit{FOXSI} DSSD \citep{ishikawa2011}, VATA450 has a higher dynamic range in energy measurements up to a soft gamma-ray range of $\approx$800~keV.  
The VATA450 has 64 readout channels and each channel has two amplifiers: one for a trigger and one for pulse height measurement.  
Since the \textit{FOXSI} CdTe detector has 128 channels on each of the HV and LV sides, 4 ASICs in total are used to read out all the channels of a single detector.  
Once a signal is triggered in any channel, all the pulse heights of all the channels are recorded by sample-and-hold circuits.  
The readout system using VATA450 can read out signals with a count rate up to 500~counts/s, 
and the requirement for the count rate tolerance is met.  

In order to achieve the fine 60~$\mu$m strip pitch, glass fanout boards to connect the electrode strips and ASICs were newly developed.  The glass boards are bump-bonded to the detector using gold/indium stud bumps with optimized temperature and pressure.  We successfully established the conditions and procedure for connecting the detector strips to the readout circuit, and two flight detectors were successfully fabricated with most of the strips connected to the readout circuits excluding a few strips.  

\begin{figure}[htbp]
 \begin{center}
 \noindent\includegraphics[width=20pc]{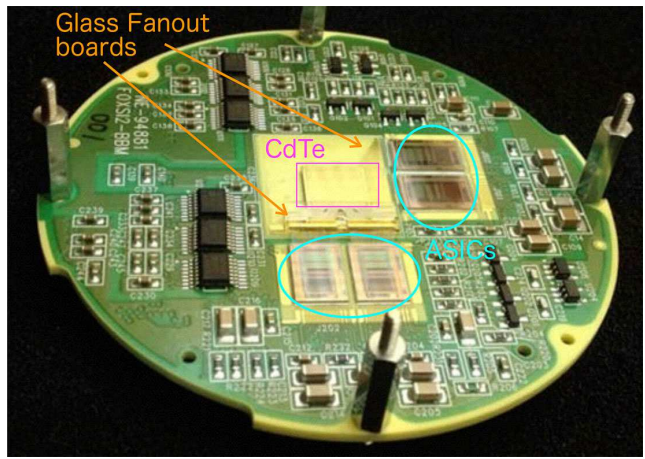}
 \caption{Photo of the \textit{FOXSI} CdTe detector board.  The CdTe detector is in the center of the board, and 4 readout ASICs are in front and to the right of the detector. }
 \label{figure_photo}
 \end{center}
 \end{figure}

A photo of the detector board is shown in Fig.~\ref{figure_photo}.  
The CdTe sensor is placed at the center on the board, and 4 ASICs are placed to read out the signals.  
Two ASICs on the right read out the strips on the bottom side, and the other two ASICs at the front side of the detector read the strips on the top side.  The glass fanout boards are located underneath the detector and (at the front side only), slightly overlapping the top of the detector.  The top side is the LV side.  
With this configuration, HXRs are incident on the LV side, and the probability of interactions near the LV surface is higher than that near the HV surface.  
Contributions of electrons are dominant for readout signals if incident photons are absorbed closer to the LV surface. 
Since the mobility lifetime product of electrons is higher than that of holes, charge loss is minimized with this configuration.  

\section{Laboratory performance measurements of the \textit{FOXSI} CdTe detector} 
\begin{figure*}[htbp]
  \begin{center}
 \noindent\includegraphics[width=38pc]{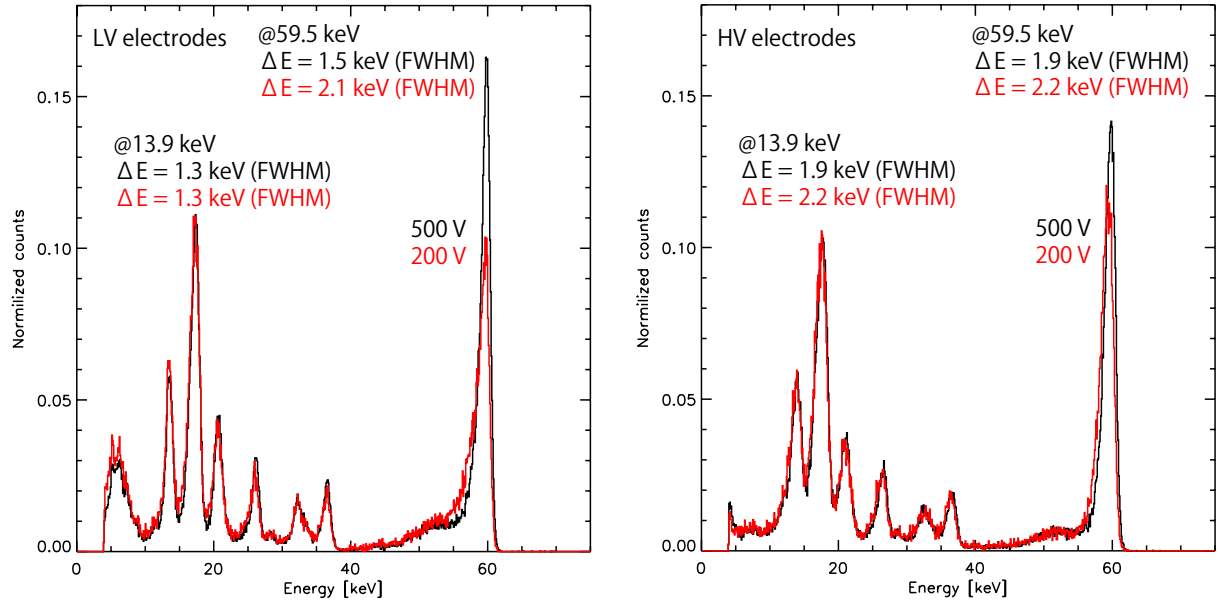}
 \caption{Single-hit spectra for a $^{241}$Am source obtained from the LV (left) and HV (right) electrodes of the \textit{FOXSI} CdTe detector.  The bias voltages were 500~V (black) or 200~V (red), and the temperature was $-20~^{\circ}$C.  
 The source was irradiated on the HV surface of the detector.}
 \label{figure_spec_1hit}
 \end{center}
 \end{figure*}

We tested the \textit{FOXSI} CdTe detector in the laboratory to assess whether the performance meets the requirements.  
The spectra of an Americium 241 ($^{241}$Am) radioisotope source for the LV (left panel) and HV (right panel) readouts are shown in Fig.~\ref{figure_spec_1hit}.  
Signals of all available strips are included, and events in which only one strip on each side has significant signal (``single hit events'') are used for these spectra.  
In general, the spectral performance is better with 0.5~mm thick CdTe detector if high bias voltages up to $\approx$500~V are applied because the charge collection efficiency is higher (due to the low mobility-lifetime products of the charge carriers).
However, the performance does not change drastically for lower energies such as the \textit{FOXSI} rocket energy range of $<$15~keV because most of the incident photons interact near the surface of the detector.  
Because of that and because of existing electronic designs from the first \textit{FOXSI} flight, the bias voltages was set to 200~V during the \textit{FOXSI}-2 flight.
We show results for both 200~V (red) and 500~V (black) bias voltages applied.  
The energy resolutions at the 13.9~keV gamma-ray line were measured to be 1.3~keV and 1.9~keV (FWHM) by the LV and HV electrodes with 500~V.
The 13.9~keV resolutions were similar or bit worse with 200~V (1.3~keV and 2.2~keV FWHM, respectively).  
A 1.3~keV resolution is good enough to measure photon energies down to 4~keV,
because 4~keV corresponds to $>$7 sigma of the energy resolution; therefore we can be confident of successful photon detection for a 4~keV signal.  
The energy resolutions at the 59.5~keV gamma-ray line were 1.5~keV FWHM for LV and 1.9~keV FWHM for HV at 500~V, and 2.1~keV FWHM for LV and 2.2~keV FWHM for HV at 200~V.

\begin{figure*} [htbp]
 \begin{center}
 \noindent\includegraphics[width=38pc]{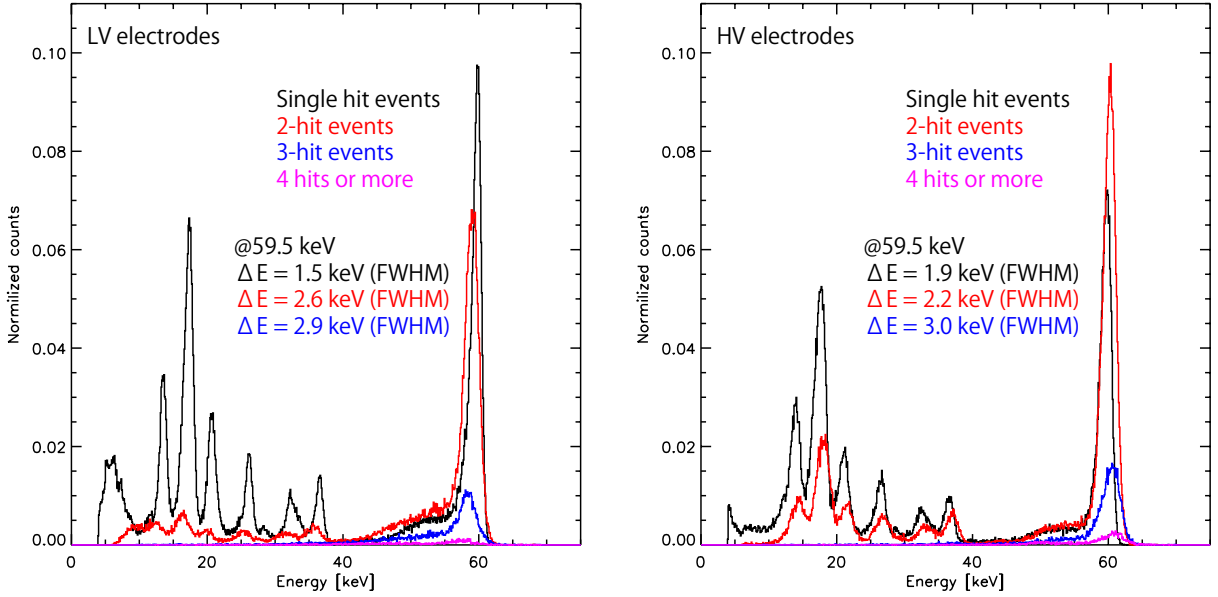}
 \caption{$^{241}$Am Spectra obtained for the LV (left) and HV (right) electrodes of the \textit{FOXSI} CdTe detector.  
 Spectra of single-hit events (black), 2-hit events (red), 3-hit events (blue) and events with 4 or more hits (magenta) are separately shown. 
 The bias voltage was 500~V (black) and the temperature was $-20~^{\circ}$C.}
 \label{figure_spec_hits}
 \end{center}
 \end{figure*}

Since the strip size is small in comparison to a typical charge cloud, a single incident photon might produce significant signals in multiple strips per side. 
Those events are called ``split events'' or ``charge-shared events.''  
In Fig.~\ref{figure_spec_hits}, the $^{241}$Am spectra for single-hit events (black) and split events (red: 2-hit events, blue: 3-hit events, magenta: events with 4 and more hits) are shown for the LV (left panel) and HV (right panel) electrodes.  
Signals from multiple strips are summed for each event, and the spectra include events from all available strips.  
The bias voltage was 500~V and the temperature was $-$20~$^{\circ}$C.
Although the energy resolutions are worse for split events, e.g. 2.6~keV (LV) and 2.2~keV (HV) FWHM for the 2-hit events at 59.5~keV, 
the information about the incident photon energy is preserved and those events can be used for spectroscopy.  
In addition, positional information finer than the strip pitch can be obtained with split events by using the ratio of the hit signals.  
 
The ratio of split events with 500~V and 200~V for the 10--30~keV lines and the 59.5~keV line are summarized in Table~1 (LV) and Table~2 (HV).
The data used are the same as in Fig.~\ref{figure_spec_hits}.  
Note that the errors shown in the tables are 1-sigma errors and statistical errors only.  
We can see that a significant fraction of the events are split events for the 59.5~keV line for both the LV and HV electrodes.  
On the other hand, the single hit events are dominant ($>$70~\%) for the 10--30~keV lines.  
With 200~V, it is clearly seen that the ratios of single hit events is smaller and split events are more numerous as compared with the 500~V case.  
However, the difference of the split events between the 500~V and 200~V cases is only a few percent.  
For the \textit{FOXSI} rocket energy range, most events are single hit events.  
\begin{table}
\caption{Fraction of split events for the LV readouts.}
\centering
\begin{tabular}{c c c c c c}
\hline  
 Energy & Bias    &  single hit & 2-hit &3-hit & 4-hit or\\
 {[keV]} & voltage &  [\%]& [\%] & [\%] & more \\
  & [V] &  &  &  &  [\% ] \\
\hline  \hline
10--30 & 500& 85.8$\pm$0.5 & 13.9$\pm$0.2& 0.3$\pm$0.02& 0.09$\pm$0.01  \\
 & 200& 82.6$\pm$0.8 & 16.7$\pm$0.3& 0.5$\pm$0.04& 0.1$\pm$0.02  \\
 \hline
59.5 & 500& 41.4$\pm$0.2 & 48.1$\pm$0.2& 9.4$\pm$0.08& 1.1$\pm$0.03  \\
 & 200& 36.0$\pm$0.3 & 51.4$\pm$0.4& 11.0$\pm$0.2& 1.6$\pm$0.06  \\
\hline 
\end{tabular}
\end{table}
\begin{table}
\caption{Fraction of split events for the HV readouts.}
\centering
\begin{tabular}{l c c c c c}
\hline  
 Energy & Bias    &  single hit & 2-hit &3-hit & 4-hit or\\
 {[keV]} & voltage &  [\%]& [\%] & [\%] & more \\
  & [V] &  &  &  &  [\% ] \\
\hline  \hline
10--30 & 500& 74.4$\pm$0.4 & 25.5$\pm$0.2& 0.1$\pm$0.01& 0.02$\pm$0.005  \\
 & 200& 72.7$\pm$0.6 & 26.7$\pm$0.3& 0.5$\pm$0.04& 0.07$\pm$0.01  \\
 \hline
59.5 & 500& 34.8$\pm$0.2 & 51.2$\pm$0.2& 12.1$\pm$0.1& 1.9$\pm$0.04  \\
 & 200& 33.2$\pm$0.3 & 51.7$\pm$0.4& 12.5$\pm$0.2& 2.6$\pm$0.08  \\
\hline  
\end{tabular}
\end{table}

\begin{figure}[htbp]
  \begin{center}
 \noindent\includegraphics[width=20pc]{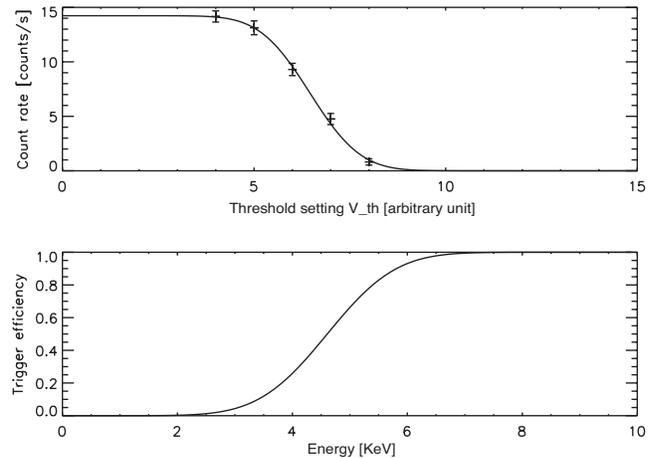}
 \caption{(Top) Plot of count rate with several trigger threshold settings using the identical condition of the radioisotope source $^{55}$Fe.
 (Bottom) Estimated trigger efficiency dependence on energies of incident photons with the flight threshold setting.  }
 \label{figure_vth}
 \end{center}
 \end{figure}

The efficiency of the detector at low energies is determined by the trigger probability as a function of energy.  
Although the trigger threshold $V_{th}$ can be set by the slow control register to the ASICs, noise triggers become too frequent if we set the threshold too low.  
We optimize the threshold by setting it as low as possible, but high enough so as to receive almost no triggers with no source present (significantly less than 1~count$/$s).  
With $V_{th} = 5$ (in arbitrary unit), the noise trigger rate was significantly less than 1~count/s.  
While the detector also work OK with $V_{th} = 4$, the noise trigger rate was higher than 1~count/s.  
We decided to use $V_{th} = 5$ for the flight by considering the risk of more noise triggers due to the environment difference between on the ground and in flight.  
To evaluate the trigger efficiency around the low energy threshold, 
we estimated the relation between the threshold energies and threshold setting values $V_{th}$, and pulse height uncertainty of the trigger circuit $\sigma _{trig}$.  
$\sigma _{trig}$ includes noise in the trigger circuit and gain variations of the readout channels, and have different values for each ASIC.
We estimated the trigger efficiency for each ASIC as follows: 
We assumed the threshold setting $V_{th}$ to be linear with the threshold energy $E_{th}$, or $E_{th} = G V_{th} $ with a constant value $G$, since the trigger threshold is set by a standard discriminator.  
We put an iron-55 ($^{55}$Fe) radioactive source close to the detector and measured the count rate of the 5.8--6.5~keV fluorescent X-ray line complex with several $V_{th}$ settings.  
An example of the results for one ASIC of a flight detector is plotted in the top panel of Fig.~\ref{figure_vth}.  
Assuming the pulse height spectrum for the trigger circuit of the 5.9~keV line has a Gaussian shape $\propto \exp [-(E-5.9{\rm \mbox{ }keV})^2 / 2\sigma _{trig} ^2]$ where $E$ is the energy, 
the count rate $C$ in the 5.9~keV line is expressed by
\begin{eqnarray}
C &=& C_0 \left[ 1  -\frac{1}{\sqrt{2\pi} \sigma_{trig}} \right. \nonumber \\
& & \left. \times \int _{-\infty}^{E_{th}} \exp \left\{ -\frac{(E-{\rm 5.9\mbox{ }keV})^2} {2\sigma _{trig} ^2} \right\} \mathrm{d}E \right] , \nonumber \\
&=& \frac{C_0}{2} \mathrm{erfc}\left( \frac{GV_{th}-{\rm 5.9\mbox{ }keV}} {\sqrt{2}\sigma _{trig} }\right) , \label{eq_rate}
\end{eqnarray}
where $C_0$ is the count rate for 100~\% trigger efficiency.
The 5.8--6.5~keV emissions from the $^{55}$Fe source consists of several lines, and we used a function 
that sums the same form of the functions as eq.~\ref{eq_rate} for the dominant 4 lines (manganese K$_{\alpha 1}$, K$_{\alpha 2}$, K$_{\beta 1}$ and K$_{\beta 3}$ lines).  
We ignored the other lines because their photon emission rates are much less than 0.1~\% compared to the dominant 4 lines. 
The curve in the top panel of Fig.~\ref{figure_vth} is the result of fitting the function with free parameters $G$, $\sigma _{trig}$ and the normalization factor.  
Using obtained parameters $G$ and $\sigma _{trig}$, the trigger efficiency $\epsilon$ at an incident photon energy $E$ can be calculated by 
\begin{eqnarray}
\epsilon (E) =  \frac{1}{2} \left[ \mathrm{erf} \left( \frac{ E-GV_{th}}{\sqrt{2} \sigma _{trig}} \right) +1 \right] .
\end{eqnarray}
The calculated trigger efficiency curve for the ASIC with the \textit{FOXSI} flight setting of $V_{th} = 5$ is plotted in the lower panel in Fig.~\ref{figure_vth}.  
The efficiency was calculated to be 26\%, 50\% and 90\% for 4, 5 and 6 keV; therefore we confirmed that the detector has appreciable sensitivity down to 4~keV.  
The Sun produces photon spectra that fall steeply with energy in the \textit{FOXSI} energy range, and 26~\% efficiency is sufficient to detect 4~keV photons with good statistics with \textit{FOXSI}.  
In summary, we have confirmed that the  \textit{FOXSI} CdTe detector meets the energy range requirements.

\section{In-flight performance of the \textit{FOXSI} CdTe detector} 
\begin{figure*}[htbp]
  \begin{center}
 \noindent\includegraphics[width=38pc]{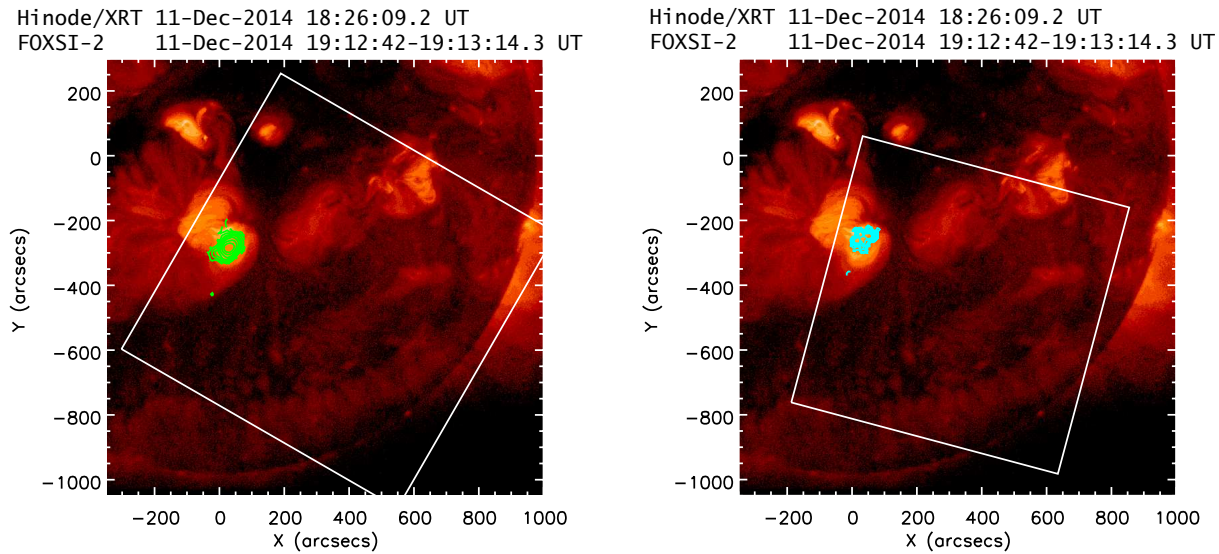}
 \caption{\textit{FOXSI} count contours for one DSSD (left) and one CdTe detector (right) plotted on the soft X-ray image taken by the X-Ray Telescope of the \textit{Hinode} satellite.  The white boxes show the fields of view of the \textit{FOXSI} detectors.  }
 \label{figure_image}
 \end{center}
 \end{figure*}

The \textit{FOXSI}-2 payload was successfully launched on December 11, 2014 from the White Sands Missile Range, New Mexico, USA.  
\textit{FOXSI}-2 has seven detectors, and we used 2 CdTe detectors and 5 DSSDs. 
\textit{FOXSI}-2 successfully imaged HXR emissions from multiple regions.  
The obtained images from the first pointing (32.3~s duration) by one silicon and one CdTe detector are shown in Fig.~\ref{figure_image}.  
Contours depict the $>4$~keV counts detected by the \textit{FOXSI} DSSD (left panel) and CdTe detector (right panel), and the background image is the soft X-ray image observed 
$\approx$46.5~minutes before the \textit{FOXSI} observation by the X-Ray Telescope of the \textit{Hinode} satellite with an Al-mesh filter. 
The white boxes show the fields of view of the \textit{FOXSI} silicon and CdTe detectors.  

An identical HXR source is detected by the Si and CdTe detectors, and the capability of the CdTe double-sided strip detector for solar HXR observations is successfully demonstrated.
Scientific results from the successful \textit{FOXSI}-2 observations will be shown soon in separate papers.  

\section{Summary and future work} 
We developed a fine-pitch CdTe double-sided strip detector for solar HXR observations with the \textit{FOXSI} sounding rocket experiment.  The fine position resolution of 60~$\mu$m and good energy resolution of 1.3~keV (FWHM at 13.9~keV) are achieved and solar HXR images were successfully obtained during the \textit{FOXSI}-2 flight.  
In \textit{FOXSI}-2, only 2 CdTe detectors were used and the other 5 detectors were DSSDs. 
The third launch opportunity of \textit{FOXSI} (\textit{FOXSI}-3) is already funded by NASA to be launched in 2018, and we plan to use the \textit{FOXSI} CdTe detector for all 7 detectors.
After \textit{FOXSI}-3, for better understanding of the HXR sources in the Sun including large solar flares, long duration observation with a wider energy band using a spacecraft is desired.  
For focusing HXR observations of the Sun by a spacecraft, it is necessary to update the readout system to handle a much higher count rate to observe large flares in addition to the size optimization.  
The concept of the CdTe double-sided strip detector for HXR observations of the Sun is already successfully demonstrated by this study, and 
we will develop detectors for a spacecraft based on the experience described in this article.


%
%
%
%
%
%
%

\begin{acknowledgments}
The development of the detectors was supported through KAKENHI Grants Number 24244021 and 20244017 from the Japan Society for the Promotion of Science.  
\textit{FOXSI} was funded by NASA's LCAS program, grant NNX11AB75G. 
\textit{Hinode} is a Japanese mission developed and launched by ISAS/JAXA, collaborating with NAOJ as a domestic partner, and NASA and STFC (UK) as international partners. Scientific operation of the \textit{Hinode} mission is conducted by the \textit{Hinode} science team organized at ISAS/JAXA. This team mainly consists of scientists from institutes in the partner countries. Support for the post-launch operation is provided by JAXA and NAOJ (Japan), STFC (U.K.), NASA, ESA, and NSC (Norway).
The authors will provide the data shown in this article to whom interested.  Please contact S.I.: s.ishikawa@solar.isas.jaxa.jp.
\end{acknowledgments}

\end{article}
%
%
%
%
%
%
%
%


\end{document}